\documentclass[twocolumn,10pt]{IEEEtran}
\usepackage[T1]{fontenc}
\usepackage{calc}
\usepackage{amsthm}
\usepackage{color}
\usepackage{amsmath}
\usepackage{newtxtext,newtxmath}
\usepackage{cite}
\usepackage{graphicx} 
\graphicspath{{../figures/}}
\usepackage[unicode=true,
 bookmarks=false,bookmarksnumbered=true,bookmarksopen=false,bookmarksopenlevel=1,
 breaklinks=false,pdfborder={0 0 0},pdfborderstyle={},backref=false,colorlinks=false]
 {hyperref}
\hypersetup{pdftitle={Title},
 pdfauthor={Admin},
 pdfpagelayout=OneColumn, pdfnewwindow=true, pdfstartview=XYZ, plainpages=false}
\makeatletter

\DeclareRobustCommand*{\lyxarrow}{%
\@ifstar
{\leavevmode\,$\triangleleft$\,\allowbreak}
{\leavevmode\,$\triangleright$\,\allowbreak}}

 \let\oldforeign@language\foreign@language
 \DeclareRobustCommand{\foreign@language}[1]{%
   \lowercase{\oldforeign@language{#1}}}

\usepackage[caption=false,font=footnotesize]{subfig}
\theoremstyle{plain}

\theoremstyle{plain}

\makeatother

\providecommand{\lemmaname}{Lemma}
\newtheorem{theorem}{Theorem}
\newtheorem{rem}{Remark}

\begin{document}

\title{On the Uplink Achievable Rate of Massive MIMO System With Low-Resolution ADC and RF Impairments}
\author{Liangyuan~Xu, Xintong~Lu, Shi~Jin, Feifei~Gao and~Yongxu~Zhu
\thanks{L.~Xu and F.~Gao  are with the Department of Automation, Tsinghua University,
Beijing 100084, China (e-mail: \protect\href{mailto:xly18@mails.tsinghua.edu.cn}{xly18@mails.tsinghua.edu.cn}; \protect\href{mailto:feifeigao@ieee.org}{feifeigao@ieee.org}).}
\thanks{X.~Lu and S.~Jin are with the National Mobile Communications Research Laboratory, Southeast University, Nanjing 210096, China (e-mail: \protect\href{mailto:luxintong@seu.edu.cn}{luxintong@seu.edu.cn}; \protect\href{mailto:jinshi@seu.edu.cn}{jinshi@seu.edu.cn}).}
\thanks{Y.~Zhu is with the Wolfson School of Mechanical, Electrical and Manufacturing Engineering, Loughborough University, Leicestershire,
LE11 3TU, UK (e-mail: \protect\href{mailto: yongxu.zhu.13@ucl.ac.uk }{yongxu.zhu.13@ucl.ac.uk}).}
}





\maketitle

\begin{NoHyper}
\begin{abstract}
This paper considers channel estimation and uplink achievable rate of the coarsely quantized massive multiple-input multiple-output (MIMO) system with radio frequency (RF) impairments. We utilize additive quantization noise model (AQNM) and extended error vector magnitude (EEVM) model to analyze the impacts of low-resolution analog-to-digital converters (ADCs) and RF impairments respectively. We show that hardware impairments cause a nonzero floor on the channel estimation error, which contraries to the conventional case with ideal hardware.   The maximal-ratio combining (MRC) technique is then used at the receiver, and an approximate tractable expression for the uplink achievable rate is derived. The simulation results illustrate the appreciable compensations between ADCs' resolution and RF impairments. The proposed studies support the feasibility of equipping economical coarse ADCs and economical imperfect RF components in practical massive MIMO systems.
\end{abstract}

\begin{IEEEkeywords}
Quantized massive MIMO, uplink rate, channel estimation,  RF impairments, low-resolution ADC, MRC.
\end{IEEEkeywords}

%
\IEEEpeerreviewmaketitle


\section{Introduction}

\IEEEPARstart{M}{assive} multi-input multi-output (MIMO), a promising technology for 5G mobile network, deploys a large number of	radio frequency (RF) chains and analog-to-digital converters (ADCs) at the base station (BS) \cite{thomas}.  As the number and quality of ADCs and RF chains increase, the financial costs and energy dissipation will grow significantly, which motivates studies of equipping economical coarse ADCs and imperfect RF chains in massive MIMO system.

Under the assumption of additive quantization noise model (AQNM), the impacts of low-resolution ADCs on the  uplink achievable rate of massive MIMO system were investigated in \cite{li,DLL}, and the asymptotic downlink achievable rate was derived in \cite{xuweiwcl}. For the special case of 1-bit quantization, channel estimation and performance of massive MIMO system have been investigated in \cite{lyz}. These studies, however, ignored RF impairments, e.g., amplifier nonlinearities, I/Q imbalance and phase noise.

On the other hand, the effects of I/Q imbalance were analyzed in \cite{IQmis}. To capture the aggregate impact of different types of RF impairments, \cite{schenkbook} proposed a generalised error model, named extended EVM (EEVM). However, low-resolution ADCs were not taken into account.

The impacts of both ADCs and RF impairments on the energy efficiency, capacity and estimation were investigated in \cite{emil}. However, the overall impacts were modeled as additive Gaussian noise which is excessively general. 

In this paper, we investigate the uplink achievable rate and channel estimation of massive MIMO system with both low-resolution ADCs and RF impairments. Instead of modeling these impacts as simple additive Gaussian noise, we utilize AQNM and EEVM model to capture the impacts of coarse ADCs and RF impairments respectively. Specifically, we first propose an approach for channel estimation under minimum mean square error (MMSE) criterion, and we demonstrate that the estimation accuracy is limited by both coarse ADCs and hardware impairments.  Then, the maximal-ratio combining (MRC) technique is applied at the receiver with imperfect channel state information (CSI), and a tightly approximated tractable expression of the uplink achievable rate is derived. We show that increasing the number of receiver antennas could mitigate the performance degradation caused by both coarse ADCs and RF impairments. In addition, the appreciable compensations between  ADCs' resolution and RF impairments are illustrated, which indicates that the performance loss caused by severe RF impairments  could be compensated by improving the resolution of ADCs, and vice versa. These compensations is valuable and could be used to optimize the financial costs and energy dissipation of massive MIMO system.

\section{System Model}
Consider a multi-user massive MIMO system consisted of a BS with $M$ antennas and $K$ single-antenna users, as demonstrated in Fig. \ref{fig:systdiagram}. Assume that RF chains and ADCs of the BS are ideal. The received signal vector at the BS is
\begin{equation}\label{eq:channelG}
{\bf{y}} = \sqrt {\rho} {{\bf{Gx}}} + {{\bf{n}}},
\end{equation}
where ${\bf G}$ is the $M\times K$ channel matrix with the $(m,k)$th element $g_{mk}$, $\mathbf{x}$ denotes the $K\times1$ symbols vector transmitted by $K$ users, ${\rho}$ is the normalized average  power of each user, and ${\bf n}\sim\mathcal{CN}(0,{\bf I})$ is the additive white Gaussian noise vector.\par
The channel coefficient between the $k$th user and the $m$th antenna of the BS is modeled as
\begin{equation}
{g_{mk}} = {h_{mk}}\sqrt {{\beta _k}},
\end{equation}
where $h_{mk}\sim \mathcal{CN}(0,1)$ is the fast-fading coefficient, and $\beta_{k}$  presents both geometric attenuation and shadow fading of the $k$th user to the whole antenna array \cite{thomas}.  

With the existence of errors caused by imperfect RF chains, we should adopt EEVM to rewrite the received signal as \cite[Chapter 7]{schenkbook}
\begin{equation} \label{eq:RFch}
{\bf{y}}_{\rm RF} =  \sqrt {{\rho}} {\boldsymbol{\chi}} {{\bf{Gx}}} + {\bf{n}}_{\rm RF} + {\bf{n}},
\end{equation}
where ${\bf{y_{\rm RF}}}$ is the received vector after imperfect RF chains, 
$
{\boldsymbol{\chi}}={\rm{diag}}\{ \chi(1),\cdots,\chi(M) \},
$
$
{\bf{n}}_{\rm RF} =\{ {n}_{\rm RF}(1),\cdots,{n}_{\rm RF}(M)\}^{T},
$ $n_{\rm RF}(m)$ denotes the additive distortion noise of the $m$th RF chains, and $\chi(m)=\kappa(m)e^{\jmath\varphi(m)}$ presents  scaling and phase shift effects of the $m$th RF chains with $\vert \kappa(m) \vert\leq 1$. The mapping of these parameters to particular type of RF impairment (e.g., nonlinearity, I/Q imbalance and phase noise) could be found in \cite[Chapter 7]{schenkbook}.   For ease of derivation, we assume that $n_{\rm RF}(m)$ is Gaussian with $n_{\rm RF}(m) \sim \mathcal{CN}(0,\sigma^2_m)$, and impairments of all RF chains are in the same level with ${ \chi(m) }= \chi $ and ${\sigma_m}=\sigma$ in the remainder of this paper.

Assuming the automatic gain control (AGC) is ideal and set properly, we can use AQNM to model the coarsely quantized outputs as \cite{li}
\begin{equation} \label{eq:all}
{\bf{y}}_{\rm q} = \eta {\bf{y}}_{\rm RF} + {\bf{n}}_{\rm q}=\eta {\sqrt{{\rho}}} {\boldsymbol{\chi}} {\bf{G}} {\bf{x}} + \eta {\bf{n}}_{\rm RF} + \eta {\bf{n}} + {\bf{n}}_{\rm q},
\end{equation}
where ${\bf{n}}_{\rm q}$ is the additive quantization noise vector such that ${\bf n_{\rm q}}$ and ${\bf y}_{\rm RF}$ are uncorrelated, $\eta=1-\mu$, and $\mu$ is the inverse of signal-to-quantization-noise ratio. We define ${\mathbf{P}}\triangleq{\boldsymbol{\chi}} {\bf{G}}$ as \emph{effective channel}. Let $b$ denotes the quantization bits. Then, $\mu$ can be approximately expressed as $\mu=\frac{\pi\sqrt{3}}{2}2^{-2b}$ for $b>5$, and the values of $\mu$ for $b\leq5$ are listed in Table~\ref{table:mu}\cite{amine}.

\begin{figure}[!tp]
\centering
\includegraphics[width=0.40\textwidth]{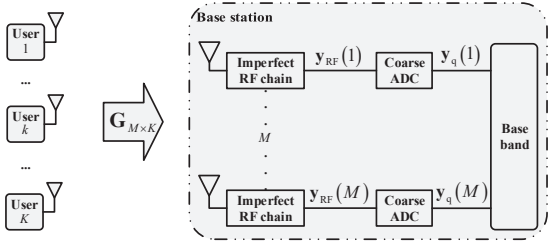}
\caption{Block diagram of multi-user massive MIMO system}
\label{fig:systdiagram}
\end{figure}
\begin{table}[!tbp]
\centering
\caption{$\mu$ for Different Quantization Bits $b$}
\label{table:mu}
\begin{tabular}{|c|c|c|c|c|c|}
\hline
$b$&1&2&3&4&5\\
\hline
$\mu$&0.3634&0.1175&0.03454&0.009497&0.002499\\
\hline
\end{tabular}
\end{table}
For given channel realizations ${\bf{G}}$, the covariance matrix of ${\bf n_{\rm q}}$ can be expressed as \cite{amine}
\begin{equation}\label{eq:nq}
{\bf{C}}_{{\bf{n}}_{\rm q}}= {\mathbb E}\!\left\{ {\bf{n}}_{\rm q} {\bf{n}}_{\rm q}^H \vert {{\bf{G}}} \right\}= {\mathbb E}\left\{ \eta \left({1 - \eta } \right) {\rm{diag}}\{  {\bf{y}}_{\rm RF} {\bf{y}}_{\rm RF}^H \} \right\}. 
\end{equation}
Assume that ${\bf{C}}_{x}$ is the  covariance matrix of input signal with ${\bf{C}}_{x}=\bf{I}$, and ${\bf{C}}_{{\bf{n}}_{\rm RF} }$ is the covariance matrix of ${\bf{n}}_{\rm RF} $ with ${\bf{C}}_{{\bf{n}}_{\rm RF} }=\sigma^2\bf{I}$ where $\sigma^2$ is variance of $n_{\rm RF}(m)$. Then, \eqref{eq:nq} can be simplified as
\begin{equation}\label{eq:covar of nq}
{\bf{C}}_{{\bf{n}}_{\rm q}}=\eta \left({1 - \eta } \right){\rm{diag}} \{ {\rho} {\boldsymbol \chi} {\bf{G}} {\bf{G}}^H {\boldsymbol \chi}^H + {\bf{I}} + {\bf{C}}_{{\bf{n}}_{\rm RF} } \}.
\end{equation}
\section{Channel Estimation}
We consider  a block fading scenario where the channel remains constant during the coherent interval. Each interval is divided into two parts: one part for pilot sequences and the other for data. During pilot sequences transmission, each user transmits $\tau$ pilot symbols simultaneously. Combining $\tau$ quantized vectors of \eqref{eq:all} into a matrix yields
\begin{equation}\label{eq:CE mtx all}
{{\mathbf{Z}}_{\rm{q}}} =\left[{\bf{y}}^{1}_{{\rm q}},\cdots,{\bf{y}}^{\tau}_{{\rm q}}\right]= \eta \sqrt {{\rho_{\rm p}}} {\mathbf{P}}{{\boldsymbol{\Phi }}^T} + \eta {{\mathbf{N}}_{{\rm{RF}}}} + \eta {\mathbf{N}} + {{\mathbf{N}}_{\rm{q}}},
\end{equation} 	 
where ${{\mathbf{Z}}_{\rm{q}}}\in {\mathbb C}^{M\times\tau}$ is the quantized outputs, ${\rho_{\rm p}}$ is the power of pilot sequences, ${{\mathbf{N}}_{\rm{q}}}$, ${{\mathbf{N}}}$ and ${{\mathbf{N}}_{\rm{RF}}}$ are matrix forms of ${{\mathbf{n}}_{\rm{q}}}$, ${{\mathbf{n}}}$ and ${{\mathbf{n}}_{\rm{RF}}}$ respectively, and ${\boldsymbol{\Phi }}\in\mathbb{C}^{\tau\times K}$ ($\tau\geq K$) denotes the pilot matrix. We take ${\boldsymbol{\Phi }}$ as $K$ columns of the $\tau\times \tau$ DFT (Discrete Fourier Transform) matrix such that ${\boldsymbol{\Phi }}$ is column-wise orthogonal. 

Let us vectorize ${{\mathbf{Z}}_{\rm{q}}}$ and obtain 
\begin{equation}\label{eq:CE all}
{{\mathbf{z}}_{\rm{q}}} = {\rm vec}({{\mathbf{Z}}_{\rm{q}}})= \eta \sqrt {{\rho_{\rm p}}} \overline{\boldsymbol{\Phi }} \underline {\mathbf{p}}  + \eta {\underline {\mathbf{n}} _{{\rm{RF}}}} + \eta \underline {\mathbf{n}}  + {\underline {\mathbf{n}} _{\rm{q}}},
\end{equation}
where $\overline {\boldsymbol{\Phi }}  = \left( {{\boldsymbol{\Phi }} \otimes {{\mathbf{I}}_M}} \right)$, $\otimes$ denotes the Kronecker product, $\underline {\mathbf{p}}={\rm{vec}}\left(  {\mathbf{P}} \right)$ is the vector form of the \emph{effective channel} ${\mathbf{P}}$, ${\underline {\mathbf{n}} _{{\rm{RF}}}}={\rm{vec}}\left(  {\mathbf{N}} _{{\rm{RF}}} \right)$, ${\underline {\mathbf{n}} }={\rm{vec}}\left(  {\mathbf{N}} \right)$ and ${\underline {\mathbf{n}} _{{\rm{q}}}}={\rm{vec}}\left(  {\mathbf{N}} _{{\rm{q}}} \right)$. 
\begin{theorem}
\label{theorem:CE}
The linear minimum mean square error (LMMSE) estimator of $\underline {\mathbf{p}}$ is \cite{SMkay}
\begin{equation}\label{eq:lmmse}
\widehat {\underline {\mathbf{p}} } = {{\mathbf{C}}_{\underline {\mathbf{p}} {{\mathbf{z}}_{\rm{q}}}}}{\mathbf{C}}_{{{\mathbf{z}}_{\rm{q}}}}^{ - 1}{{\mathbf{z}}_{\rm{q}}},
\end{equation}
where ${\mathbf{C}}_{\underline {\mathbf{p}} {{\mathbf{z}}_{\rm{q}}}}$ is the covariance matrix between $\underline {\mathbf{p}}$ and ${{\mathbf{z}}_{\rm{q}}}$, ${\mathbf{C}}_{{{\mathbf{z}}_{\rm{q}}}}$ is the covariance matrix of ${{\mathbf{z}}_{\rm{q}}}$, and $\widehat {\underline {\mathbf{p}} }={\rm vec}(\widehat { {\mathbf{P}} })$ is the estimator of the  effective channel. The normalized MSE is
\begin{equation}\label{eq:MSE}
{\rm{MSE}} = \frac{\mathbb{E}\left\{ {\left\| {\widehat {\underline {\mathbf{p}} } - \underline {\mathbf{p}} } \right\|_2^2} \right\}}{{MK}} = \frac{ {\sum\limits_{k = 1}^K \left( {{\beta _k}{{\left| {{\chi}} \right|}^2} - {\alpha _{k}}{\beta _k}{{\left| {{\chi}} \right|}^2}} \right) } }{{K}},
\end{equation}
where
\begin{equation}\label{eq:alpha}
{\alpha _{k}} \triangleq \frac{{\eta {\rho _{\rm{p}}}\tau {{\left| \chi  \right|}^2}{\beta _k}}}{{\eta {\rho _{\rm{p}}}\tau {{\left| \chi  \right|}^2}{\beta _k} + (1 - \eta ){\rho _{\rm{p}}}{{\left| \chi  \right|}^2}\sum\limits_{k = 1}^K {{\beta _k} + } {\sigma ^2} + 1}}.
\end{equation}
\end{theorem}
\begin{IEEEproof}
See Appendix \ref{app:1}.
\end{IEEEproof}

Note that ${\alpha _{k}}$ is interpreted as the accuracy of the estimator and is characterized by the level of hardware impairments, pilot power and pilot length. Since the denominator of \eqref{eq:alpha} is greater than the numerator, we have $0\leq{\alpha _{k}}\leq1$.  When ${\alpha _{k}}=1$, MSE in \eqref{eq:MSE} becomes zero, which means perfect CSI without estimation error. On the other hand, ${\alpha _{k}}=0$ means the worst estimator.
\begin{rem}\label{rem:1}
In the high SNR regime, if $\tau=K$, we have 
\begin{equation}
\mathop{\lim }\limits_{{\rho _{\rm{p}}} \to \infty } {\alpha_k} =\frac{\eta K \beta_k}{\eta K \beta_k+(1-\eta){\sum\limits_i^K { {{\beta _i}} } }}<1,
\end{equation}
\begin{equation}\label{eq:MSE corollary 1}
\mathop{\lim }\limits_{{\rho _{\rm{p}}} \to \infty } {\rm{MSE}} = \sum\limits_{k = 1}^K {\left( {\left( {\frac{1}{K} -  \frac{{{\eta \beta _k}}}{\eta K\beta _k+(1-\eta){\sum\limits_i^K { {{\beta _i}} } }}} \right){\beta _k}{{\left| \chi  \right|}^2}} \right)}.
\end{equation}
\end{rem}
\emph{Remark \ref{rem:1}} indicates that there is a nonzero error floor as ${\rho _{\rm{p}}} \to \infty $ which contraries to the ideal hardware case. This nonzero error floor is characterized by the level of hardware impairments and cannot be eliminated by increasing SNR. 
\section{Uplink Achievable Rate}
By using MRC technique with imperfect CSI obtained from \eqref{eq:lmmse}, we can modify the quantized signal vector of \eqref{eq:all} into
\begin{equation}\label{eq:MRC}
{\mathbf{r}} = {{{\mathbf{\widehat P}}}^H}{{\mathbf{y}}_{\rm{q}}}.
\end{equation}
Substituting \eqref{eq:all} into \eqref{eq:MRC}, we obtain
\begin{equation}
{\mathbf{r}} =  \eta \sqrt {{\rho_{\rm u}}} {{{\mathbf{\widehat P}}}^H}{\mathbf{Px}} + \eta {{{\mathbf{\widehat P}}}^H}({{\mathbf{n}}_{{\rm{RF}}}} + {\mathbf{n}}) + {{{\mathbf{\widehat P}}}^H}{{\mathbf{n}}_{\rm{q}}}.
\end{equation}
The $n$th element of ${\bf r}$ can be expressed  as
\begin{equation}
\!\!r_n\!=\!\eta \sqrt {{\rho_{\rm u}}} \widehat {\mathbf{p}}_n^H{{\mathbf{p}}_n}{x_n} \!+ \underbrace{\eta \widehat {\mathbf{p}}_n^H({{\mathbf{n}}_{{\rm{RF}}}} \!+\! {\mathbf{n}})\!+\! \widehat {\mathbf{p}}_n^H{{\mathbf{n}}_{\rm{q}}}\!+\! \eta \sqrt {{\rho_{\rm u}}} \!\!\!\sum\limits_{k = 1,k \ne n}^K \!\!{\widehat {\mathbf{p}}_n^H{{\mathbf{p}}_k}} {x_k}}_{\triangleq \xi}, \notag
\end{equation}
where ${{\bf{p}}}_n $ is the $n$th column of $\bf{P}$, $\widehat {\mathbf{p}}_n$ is the $n$th column of ${{\mathbf{\widehat P}}}$, and the random variable $\xi$ presents noise-plus-interference with zero mean and variance 
\begin{equation}\label{eq:MRC I}
\!\!\!{\cal I}_{\bf{G}} \!=\!{\eta ^2}\widehat {\bf{p}}_n^H{{\bf{C}}_{{{\bf{n}}_{{\rm{RF}}}}}}{\widehat {\bf{p}}_n} + {\eta ^2}\left\| {{{\widehat {\bf{p}}}_n}} \right\|_2^2 + \widehat {\bf{p}}_n^H{{\bf{C}}_{{{\bf{n}}_{\rm{q}}}}}{\widehat {\bf{p}}_n}+{\eta ^2}{\rho _{\rm{u}}}\!\sum\limits_{\substack{{k = 1}\\{k \ne n}}}^K \!{{{\left| {\widehat {\bf{p}}_n^H{{\bf{p}}_k}} \right|}^2}}.
\end{equation}
We model $\xi$ as additive Gaussian noise which is  uncorrelated with $x_n$. Then, we can derive the ergodic uplink achievable rate of the $n$th user as
\begin{equation}\label{eq:raw}
{R_n} = \mathbb{E}\left\{ {{{\log }_2}\left( {1 + \frac{{{\rho_{\rm u}}{\eta ^2}{{\left| {\widehat {\mathbf{p}}_n^H{{\mathbf{p}}_n}} \right|}^2}}}{{{\mathcal{I}_{\mathbf{G}}}}}} \right)} \right\},
\end{equation}
where the expectation is taken with respect to ${h_{mn}}$. Since we cannot directly derive a tractable expression from \eqref{eq:raw}, an approximate expression is presented as follows
\begin{theorem}\label{theorem:rate}
The ergodic uplink achievable rate ${R_n}$ of the ${\rm n}$th user can be approximated as 
\begin{equation}\label{eq:appxi}
\widetilde{R}_n = {\log _2}{\left({1 + \frac{{\eta}{\rho_{\rm u}}{{\left| \chi  \right|}^2}{\beta _n}\left({{\alpha_n}M + 1} \right)}{\cal D}} \right)},
\end{equation}
where ${\cal D}$ is given by
\[
{\cal D}={\rho _{\rm{u}}}{\left| \chi  \right|^2}\left( {\sum\limits_i^K {\left( {{\beta _i}} \right)}  - \eta {\beta _n}} \right) + (1 - \eta ){\alpha _n}{\rho _{\rm{u}}}{\left| \chi  \right|^2}{\beta _n} + {\sigma ^2} + 1.
\]
\end{theorem}
\begin{IEEEproof}
See Appendix \ref{app:2}.
\end{IEEEproof}

\emph{Theorem \ref{theorem:rate}} shows the impacts of coarse ADCs, RF impairments and channel estimation errors on the achievable rate. Compared to the related works in \cite{emil,li,lyz}, we consider more general case with both coarse ADC and RF impairments included. Since the expression in \emph{Theorem \ref{theorem:rate}} is complicated, the compensations between  ADCs' resolution and RF impairments are  implicit. Compensations mean that the performance loss caused by severe RF impairments  could be compensated by improving the resolution of ADCs, and vice versa. To gain insights into the compensations, we will investigate the following special cases of \emph{Theorem \ref{theorem:rate}}.
\begin{rem}\label{rem:2}
Assuming perfect CSI ($\alpha_k=1$), the upper bound of $\widetilde{R}_n$ is
\begin{equation}\label{eq:rate corollary 1}
\widetilde{R}_{n,{\rm upper}} = {\log _2}{\left({1 + \frac{{{\beta _n}+M{\beta _n}}}{ {\frac{1}{\eta}}{\sum\limits_{i = 1}^K {({\beta _i})}}+\left({\frac{1}{\eta}}-2\right){\beta _n} + {\frac{1+{\sigma}^2}{\eta \rho_{\rm u} {\vert \chi \vert}^2}}}} \right)}.
\end{equation} 
\end{rem}
If RF components are ideal and only low-resolution ADCs are considered, e.g., ${\vert \chi \vert}=1$ and ${\sigma^2}=0$,  \eqref{eq:rate corollary 1} is consistent with the result in  \cite{li}. Note that $\eta$, $\sigma^2$, $\chi$ and  $p_u$ merely appear in the denominator of \eqref{eq:rate corollary 1}, and $M$ only appears in the numerator. Therefore, it is easy to figure out that the loss of the uplink achievable rate  caused by hardware impairments could always be compensated by increasing the number of antennas $M$. The compensation by increasing $p_u$, however, is unsatisfying since $p_u$ merely appears in the term ${(1+{\sigma}^2)}/{\eta p_u {\vert \chi \vert}^2}$.  As $p_u \to \infty$, ${(1+{\sigma}^2)}/{\eta p_u {\vert \chi \vert}^2}$ will converge to zero and \eqref{eq:rate corollary 1} will converge as well. The reason is that interferences among users deteriorate as $p_u$ increases.

\begin{rem}\label{rem:3}
The approximated achievable rate in \eqref{eq:appxi} can be simplified to
\begin{equation}\label{eq:rate corollary 2}
\widetilde{R}_n = {\log _2}{\left({1 + \frac{{{\beta _n}(\alpha_n M+1)}}{ {\frac{1}{\eta}}{\sum\limits_{i = 1}^K {{\beta _i}}}+{\frac{\alpha_n \beta_n}{\eta}} -(1+\alpha_n)\beta_n + {\frac{1+{\sigma}^2}{\eta \rho_{\rm u} {\vert \chi \vert}^2}}}} \right)}.
\end{equation}
\end{rem}
Note that in the denominator of \eqref{eq:rate corollary 2}, the impacts of low-resolution ADCs and RF impairments mainly occur in the term $\frac{1+{\sigma}^2}{\eta p_u {\vert \chi \vert}^2}$,  which  unveils the compensations between resolution of ADCs and RF impairments. 
Increasing $\eta$ and decreasing $\vert \chi \vert$ (alternatively, increasing $\eta$ and increasing $\sigma$) could keep the term $\frac{1+{\sigma}^2}{\eta p_u {\vert \chi \vert}^2}$ unchanged, and vice versa. This means  the uplink rate performance degradation caused by severe RF impairments could be compensated by improving the resolution of ADCs, and vice versa. Furthermore, as mentioned in \emph{Remark \ref{rem:2}}, increasing $M$ to compensate for the uplink rate loss caused by both coarse ADCs and RF impairments is also valid here. 
 
Applying these compensations in system optimization, we can get different system setups which lead to the same performance, and then we could choose the most economical and efficient one.
\section{Numerical Results}
In this simulation, we consider a cell with radius of 900 meters, where the $K$ users are randomly and uniformly distributed excepting a central circle of the BS with radius $r_h$. The geometric attenuation and shadow fading  are defined as $\beta_{k}=z_{k}/(r_{k}/r_{h})^{v}$, where $z_{k}$ is a log-normal variable with $10\log_{10}(z_{k}) \sim {\mathcal{N}}(0,\sigma^2_{shadow})$ \cite{thomas}, and $r_{k}$ is the distance between the $k$th user and the BS. We define the uplink sum rate of the entire system as $R=\sum_{n=1}^{K}R_{n}$. The simulation parameters are listed in Table \ref{table:para}.

Fig. \ref{fig:MSEVSpower} shows MSE of the channel estimator versus SNR with different levels of hardware impairments. We can see that coarse ADCs and hardware impairments create a floor on MSE. As opposed to the case of ideal hardware, an nonzero estimation error floor arises due to hardware impairments and cannot be eliminated by increasing SNR, which is discussed in \emph{Remark \ref{rem:1}}.

Fig. \ref{fig:app} shows the approximate result in \emph{Theorem \ref{theorem:rate}} and the ergodic rate in \eqref{eq:raw} versus $M$. Since the errors between the Monte-Carlo simulation of \eqref{eq:raw} and the approximate analytical uplink rate are  negligible, the accuracy of the approximate expression in \emph{Theorem \ref{theorem:rate}}  is validated. Furthermore, we can see that the channel estimation errors cause notable loss of sum rate. Moreover, compared with the case of perfect hardware, low-resolution ADCs and RF impairments cause severe performance degradation. 

Fig. \ref{fig:compesatescaling} shows the uplink sum rate versus $M$. We can see that different levels of hardware impairments lead to the same sum rate, which illustrates a type of compensation between coarse ADCs and imperfect RF components for the performance degradation. This compensation could be described as that  the uplink rate performance degradation caused by severe RF impairments (decreasing $\vert \chi \vert$) could be compensated by increasing the resolution of ADCs, and vice versa. 

\begin{figure}[!tp]
\centering
\includegraphics[width=0.45\textwidth]{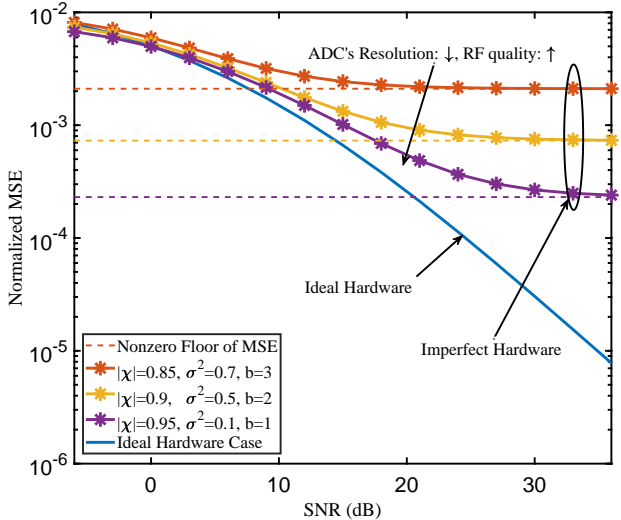}
\caption{MSE of channel estimator versus SNR.}
\label{fig:MSEVSpower}
\end{figure}
\begin{figure}[!tp]
\centering
\includegraphics[width=0.45\textwidth]{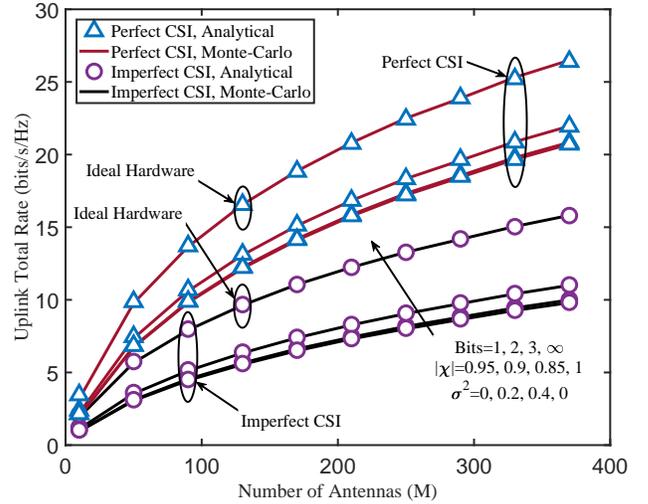}
\caption{Comparison between simulated result and analytical result.}
\label{fig:app}
\end{figure}
\begin{figure}[!tp]
\centering
\includegraphics[width=0.45\textwidth]{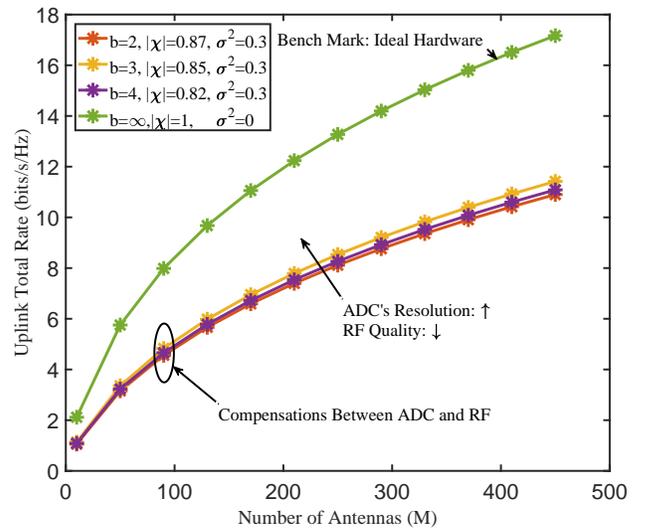}
\caption{Compensation between ADCs resolution and RF impairments.}
\label{fig:compesatescaling}
\end{figure}
\begin{table}[!tbp]
\centering
\caption{value of parameters for simulation}
\label{table:para}
\begin{tabular}{|c|c|c|c|c|c|}
\hline
parameters&$r_{h}$ (meters)&$\sigma^2_{shadow}$ (dB)&$v$&$N$&$p_u$ (dB)\\
\hline
value&100&8&3.8&10&10\\
\hline
\end{tabular}
\end{table}

\section{Conclusion}
We propose a method for channel estimation and derive a tractable expression for the uplink achievable rate of the coarsely quantized massive MIMO system with RF impairments. We show that hardware impairments and coarse ADCs create an nonzero floor on channel estimation error. Furthermore, the appreciable compensations between ADCs' resolution and RF impairments are demonstrated. These discussions support the feasibility of the deployment of coarse ADCs and imperfect RF components in massive MIMO system.

\appendices{}
\section{Proof of Theorem \ref{theorem:CE}}\label{app:1}
\begin{proof}
According to \eqref{eq:channelG}, $\bf G$ can be written as ${\bf G}={\bf H}{\bf D}^{1/2}$, where $[{\bf H}]_{m,k}=h_{mk}$, and $\bf D$ is diagonal matrix with diagonal entries $\{\beta_{k}\}$.  Effective channel then can be rewritten as ${\bf{P}} ={\boldsymbol{\chi}} {\bf{G}}={\boldsymbol{\chi}}{\bf H}{\bf D}^{1/2}$, and $\underline {\mathbf{p}}$ can be written as
\begin{equation}
\underline {\mathbf{p}}={\rm vec}\left({\boldsymbol{\chi}}{\bf H}{\bf D}^{1/2}\right)=\left({\bf D}^{1/2}\otimes{\boldsymbol{\chi}}\right){\rm vec}\left({\bf H}\right).
\end{equation}
The covariance matrix of ${\underline {\mathbf{p}} }$ is 	
\begin{equation}\label{eq:Rp}
{{\mathbf{C}}_{\underline {\mathbf{p}} }} = \mathbb{E}\left\{\underline {\mathbf{p}}\underline {\mathbf{p}}^{H}\right\}={\mathbf{D}} \otimes {\boldsymbol{\chi }}{{\boldsymbol{\chi }}^H}={{{\left| \chi  \right|}^2}}{\mathbf{D}} \otimes{\bf I}_{M},
\end{equation}
where ${\bf I}_{M}$ denotes a $M\times M$ identity matrix. According to \eqref{eq:CE all}, ${{\mathbf{z}}_{\rm{q}}} = \eta {\underline {\mathbf{z}} _{{\rm{RF}}}} + {\underline {\mathbf{n}} _{\rm{q}}}$, where ${\underline {\mathbf{z}} _{{\rm{RF}}}}={\rm vec}(\mathbf{Z} _{{\rm{RF}}})$. The covariance matrix of  ${{{\mathbf{z}}_{\rm{q}}}}$ is 
\begin{equation}\label{eq:C zq temp}
{{\mathbf{C}}_{{{\mathbf{z}}_{\rm{q}}}}} = {\eta ^2}{{\mathbf{C}}_{{{\underline {\mathbf{z}} }_{{\rm{RF}}}}}} + {{\mathbf{C}}_{{{\underline {\mathbf{n}} }_{\rm{q}}}}}={\eta ^2}{{\mathbf{C}}_{{{\underline {\mathbf{z}} }_{{\rm{RF}}}}}} + \eta (1 - \eta ){\rm{diag}}\left\{ {{{\mathbf{C}}_{{{\underline {\mathbf{z}} }_{{\rm{RF}}}}}}} \right\}.
\end{equation}
According to \eqref{eq:CE all}, we write the covariance matrix of ${{{\underline {\mathbf{z}} }_{{\rm{RF}}}}}$ as 
\begin{equation}\label{eq:C zRF temp}
{{\mathbf{C}}_{{{\underline {\mathbf{z}} }_{{\rm{RF}}}}}} = {\rho _{\rm{p}}}\overline {\boldsymbol{\Phi }} {{\mathbf{C}}_{\underline {\mathbf{p}} }}{\overline {\boldsymbol{\Phi }} ^H} + {{\mathbf{C}}_{{{\underline {\mathbf{n}} }_{{\rm{RF}}}}}} + {{\mathbf{C}}_{\underline {\mathbf{n}} }}.
\end{equation}
Substituting  $\overline {\boldsymbol{\Phi }}  = \left( {{\boldsymbol{\Phi }} \otimes {{\mathbf{I}}_M}} \right)$, ${{\mathbf{C}}_{{{\underline {\mathbf{n}} }_{{\rm{RF}}}}}}={\sigma ^2}{{\mathbf{I}}_{M\tau }}$, ${{\mathbf{C}}_{\underline {\mathbf{n}} }}={{\mathbf{I}}_{M\tau }}$ and \eqref{eq:Rp} into \eqref{eq:C zRF temp}, then we get
\begin{equation}\label{eq:Rzrf}
{{\mathbf{C}}_{{{\underline {\mathbf{z}} }_{{\rm{RF}}}}}} ={\rho _{\rm{p}}}{\left| \chi  \right|^2}{\boldsymbol{\Phi }}{\bf D}{{\boldsymbol{\Phi }}^H} \otimes {{\mathbf{I}}_M} + (1 + {\sigma ^2}){{\mathbf{I}}_{M\tau }},
\end{equation}
\begin{equation}\label{eq:diagRzrf}
{\rm{diag}}\left\{ {{{\mathbf{C}}_{{{\underline {\mathbf{z}} }_{{\rm{RF}}}}}}} \right\} = \left( {{\rho _{\rm{p}}}{{\left| \chi  \right|}^2}\sum\limits_{k = 1}^K {{\beta _k}}  + {\sigma ^2} + 1} \right){\bf I}_{M\tau}.
\end{equation}
Substituting \eqref{eq:Rzrf} and \eqref{eq:diagRzrf} into \eqref{eq:C zq temp}, we get 
\begin{equation}\label{eq:Czq}
\!\!\!\!{{\mathbf{C}}_{{{\mathbf{z}}_{\rm{q}}}}} \!=\! {\eta ^2}{\rho _{\rm{p}}}\overline {\boldsymbol{\Phi }} {{\mathbf{C}}_{\underline {\mathbf{p}} }}{\overline {\boldsymbol{\Phi }} ^H} \!+ \eta \left(\! {(1 - \eta ){\rho _{\rm{p}}}{{\left| \chi  \right|}^2}\sum\limits_{k = 1}^K {{\beta _k} + } {\sigma ^2} + 1} \!\right){{\mathbf{I}}_{M\tau }}.
\end{equation}
According to \eqref{eq:CE all}, the covariance matrix between ${\underline {\mathbf{p}} }$ and ${ {{\mathbf{z}}_{\rm{q}}}}$ is 
\begin{equation}\label{eq:C pzq}
{{\mathbf{C}}_{\underline {\mathbf{p}} {{\mathbf{z}}_{\rm{q}}}}} = \eta {{\mathbf{C}}_{\underline {\mathbf{p}} {{\underline {\mathbf{z}} }_{{\rm{RF}}}}}} + {{\mathbf{C}}_{\underline {\mathbf{p}} {{\underline {\mathbf{n}} }_{\rm{q}}}}} = \eta {{\mathbf{C}}_{\underline {\mathbf{p}} {{\underline {\mathbf{z}} }_{{\rm{RF}}}}}} = \eta \sqrt {{\rho _{\rm{p}}}} {{\mathbf{C}}_{\underline {\mathbf{p}} }}{\overline {\boldsymbol{\Phi }} ^H}.
\end{equation}
According to \eqref{eq:lmmse}, we can get 
\begin{equation}\label{eq:Cp temp}
{{\mathbf{C}}_{\widehat {\underline {\mathbf{p}} }}} = {{\mathbf{C}}_{\widehat {\underline {\mathbf{p}} }\underline {\mathbf{p}} }} = {{\mathbf{C}}_{\underline {\mathbf{p}} \widehat {\underline {\mathbf{p}} }}} ={{\mathbf{C}}_{\underline {\mathbf{p}} {{\mathbf{z}}_{\rm{q}}}}}{\mathbf{C}}_{{{\mathbf{z}}_{\rm{q}}}}^{ - 1}{{\mathbf{C}}^{H}_{\underline {\mathbf{p}} {{\mathbf{z}}_{\rm{q}}}}}.
\end{equation}
Substituting \eqref{eq:Czq} and \eqref{eq:Czq} into \eqref{eq:Cp temp}, we get 
\begin{equation}\label{eq:Cp}
{{\mathbf{C}}_{\widehat {\underline {\mathbf{p}} }}} = {{\mathbf{C}}_{\widehat {\underline {\mathbf{p}} }\underline {\mathbf{p}} }} = {{\mathbf{C}}_{\underline {\mathbf{p}} \widehat {\underline {\mathbf{p}} }}} = {\left| \chi  \right|^2}{\mathbf{D}}{\boldsymbol \alpha} \otimes {{\mathbf{I}}_M},
\end{equation}
where the matrix inverse identity $(\bf I+AB)^{-1}A=A(I+BA)^{-1}$ is applied in the derivations, and $\boldsymbol{\alpha }$ is a diagonal matrix and is given by 
\begin{equation}\label{eq:alphaMTX}
\boldsymbol{\alpha } \triangleq \frac{{\eta {\rho _{\rm{p}}}\tau {{\left| \chi  \right|}^2}{\mathbf{D}}}}{{\eta {\rho _{\rm{p}}}\tau {{\left| \chi  \right|}^2}{\mathbf{D}} + \left( {(1 - \eta ){\rho _{\rm{p}}}{{\left| \chi  \right|}^2}\sum\limits_{k = 1}^K {{\beta _k} + } {\sigma ^2} + 1} \right){{\mathbf{I}}_K}}},
\end{equation}
where the division of matrix means $\bf \frac{A}{B}=AB^{-1}$. The $k$th diagonal element of $\boldsymbol{\alpha }$ is ${\alpha _k} = {\left[ {\boldsymbol{\alpha }} \right]_{k,k}}$  with
\begin{equation}
{\alpha _k} = \frac{{\eta {\rho _{\rm{p}}}\tau {{\left| \chi  \right|}^2}{\beta _k}}}{{\eta {\rho _{\rm{p}}}\tau {{\left| \chi  \right|}^2}{\beta _k} + (1 - \eta ){\rho _{\rm{p}}}{{\left| \chi  \right|}^2}\sum\limits_{k = 1}^K {{\beta _k} + } {\sigma ^2} + 1}}.
\end{equation}
Then, the normalized MSE is given by
\begin{equation}\label{eq:MSE temp}
{\rm{MSE}} = \frac{{{\mathbb E}\left\{ {\left\| {\widehat {\underline {\bf{p}} } - \underline {\bf{p}} } \right\|_2^2} \right\}}}{{MK}} = \frac{{{\rm{tr}}\left( {{{\bf{C}}_{\widehat {\underline {\bf{p}} }}}{\rm{ + }}{{\bf{C}}_{\underline {\bf{p}} }}{\rm{ - }}{{\bf{C}}_{\widehat {\underline {\bf{p}} }\underline {\bf{p}} }}{\rm{ - }}{{\bf{C}}_{\underline {\bf{p}} \widehat {\underline {\bf{p}} }}}} \right)}}{{MK}}.
\end{equation}
Substituting \eqref{eq:Rp} and \eqref{eq:Cp} into \eqref{eq:MSE temp}, we then get
\begin{equation}
{\rm{MSE}} =  \frac{ {\sum\limits_{k = 1}^K \left( {{\beta _k}{{\left| {{\chi}} \right|}^2} - {\alpha _{k}}{\beta _k}{{\left| {{\chi}} \right|}^2}} \right) } }{{K}}.
\end{equation}
\end{proof}
\section{Proof of Theorem \ref{theorem:rate}}\label{app:2}
\begin{proof}
We assume $\underline{\boldsymbol{\delta }} = \underline {\mathbf{p}}  - \widehat {\underline {\mathbf{p}} }$ where $\underline{\boldsymbol{\delta }}$ denotes the channel estimation error vector. According to the orthogonality principle of MMSE estimator \cite{SMkay}, we can get
\begin{equation}\label{eq:orthog}
\mathbb{E}\left\{ {\left( {\underline {\mathbf{p}}  - \widehat {\underline {\mathbf{p}} }} \right){\mathbf{z}}_{\rm{q}}^H} \right\} = {\mathbf{0}},
\end{equation}
which indicates $\mathbb{E}\left\{ \underline{\boldsymbol{\delta }}\widehat{\underline {\mathbf{p}} }^{H} \right\} = {\mathbf{0}}$. Since $\underline{\boldsymbol{\delta }}$, $\underline {\mathbf{p}}$, ${\mathbf{z}} _{\rm{q}}$ and $\widehat {\underline {\mathbf{p}} }$ all are jointly Gaussian distributed, $\underline{\boldsymbol{\delta }}$ and $\widehat {\underline {\mathbf{p}} }$ are independent. To make the following derivations more clear, we rewrite $\widehat {\underline {\mathbf{p}} }$ as
\begin{equation}
\widehat {\underline {\mathbf{p}} } = {\left\{ {{{\widehat {\mathbf{p}}}_1}^T, \cdots,{{\widehat {\mathbf{p}}}_k}^T ,\cdots,{{\widehat {\mathbf{p}}}_K}^T} \right\}^T},
\end{equation}  
where ${\widehat {\mathbf{p}}_k}$ is the $k$th column of ${{\mathbf{\widehat P}}}$. Similarly we have 
\begin{equation}
\underline {\mathbf{p}}= {\left\{ {{\mathbf{p}}_1^T, \cdots, {\mathbf{p}}_k^T,\cdots,{\mathbf{p}}_K^T} \right\}^T},
\end{equation}
\begin{equation}
\underline{\boldsymbol{\delta }}={\left\{ {{\boldsymbol{\delta }}_{1}^{T}, \cdots,{\boldsymbol{\delta }}_k^T ,\cdots,{\boldsymbol{\delta }}_K^T} \right\}^T},
\end{equation}
where ${\boldsymbol{\delta }}=\mathbf{P}-\widehat {\underline {\mathbf{P}} }$ denotes error matrix,  ${\mathbf{p}}_k$ is the $k$th column of ${{\mathbf{P}}}$,  and ${\boldsymbol{\delta }}_k$ is the $k$th column of ${\boldsymbol{\delta }}$. According to \eqref{eq:Rp} and \eqref{eq:Cp}, we have ${{\bf{C}}_{\underline {\boldsymbol{\delta }} }}{\rm{ = }}{{\bf{C}}_{\underline {\bf{p}} }}{\rm{ - }}{{\bf{C}}_{\widehat {\underline {\bf{p}} }}}$. Then the covariance matrix of ${\widehat {\mathbf{p}}_k}$ can be written as  ${{\mathbf{C}}_{{{\widehat {\mathbf{p}}}_{k}}}} ={\left| \chi  \right|^2}\beta_k \alpha_k {{\mathbf{I}}_M}$, and similarly we have ${{\mathbf{C}}_{{{ {\mathbf{p}}}_{k}}}}={\left| \chi  \right|^2}\beta_k{{\mathbf{I}}_M}$ and ${{\mathbf{C}}_{{{\boldsymbol{\delta }}_k}}} =({\left| \chi  \right|^2}\beta_k -{\left| \chi  \right|^2}\beta_k\alpha_k){{\mathbf{I}}_M}$. Thus, the distributions of the $m$th elements of ${\widehat {\mathbf{p}}_k}$, ${\mathbf{p}}_k$ and ${\boldsymbol{\delta }}_k$ are 
\begin{equation}\label{eq:dis hatp}
\!\!\!\![{{\widehat {\mathbf{p}}}_k}]_{m}\sim\mathcal{CN}\left(0,{\left| \chi  \right|^2}\beta_k \alpha_k\right), \!\quad\! {\left| {{{[{{\widehat {\mathbf{p}}}_k}]}_m}} \right|^2}\sim\Gamma \left(1,{\left| \chi  \right|^2}\beta_k \alpha_k\right),
\end{equation}
\begin{equation}\label{eq:dis p}
[{{ {\mathbf{p}}}_k}]_{m}\sim\mathcal{CN}\left(0,{\left| \chi  \right|^2}\beta_k \right), \quad {\left| {{{[{{ {\mathbf{p}}}_k}]}_m}} \right|^2}\sim\Gamma \left(1,{\left| \chi  \right|^2}\beta_k \right),
\end{equation}
\begin{equation}\label{eq:dis delta}
[{{{\boldsymbol{\delta }}_k}}]_{m}\!\sim\!\mathcal{CN}\!\left( {0,{{\left| \chi  \right|}^2}{\beta _k}(1\!-\!{\alpha _k})} \right),{\left| [{{{\boldsymbol{\delta }}_k}}]_{m}\right|}^2\!\sim\!\Gamma \!\left( {1,{{\left| \chi  \right|}^2}{\beta _k}(1 \!-\! {\alpha _k})} \right),\notag
\end{equation}
where $\Gamma$ denotes Gamma distribution.

Next, we will derive the expression of $\widetilde{R}_n $. According to \cite[Lemma 1]{qi}, $R_n$ can be precisely approximated by
\begin{equation}\label{eq:rate approx temp}
{R_n} \approx\widetilde{R}_n= {\log _2}\left( {1 + \frac{{{\rho_{\rm u}}{\eta ^2}\mathbb{E}\left\{ {{{\left| {\widehat {\mathbf{p}}_n^H{{\mathbf{p}}_n}} \right|}^2}} \right\}}}{{\mathbb{E}\left\{ {{\mathcal{I}_{\mathbf{G}}}} \right\}}}} \right),
\end{equation}
where
\begin{equation}\label{eq:D temp}
\begin{split}
\mathbb{E}\left\{ {{\mathcal{I}_{\mathbf{G}}}} \right\} =& {\eta ^2}{\rho _{\rm{u}}}\mathbb{E}\left\{ {\sum\limits_{k = 1,k \ne n}^K {{{\left| {\widehat {\mathbf{p}}_n^H{{\mathbf{p}}_k}} \right|}^2}} } \right\} + {\eta ^2}\mathbb{E}\left\{ {\widehat {\mathbf{p}}_n^H{{\mathbf{C}}_{{{\mathbf{n}}_{{\rm{RF}}}}}}{{\widehat {\mathbf{p}}}_n}} \right\} \\
&+ {\eta ^2}\mathbb{E}\left\{ {{{\left\| {{{\widehat {\mathbf{p}}}_n}} \right\|}^2}} \right\} + \mathbb{E}\left\{ {\widehat {\mathbf{p}}_n^H{{\mathbf{C}}_{{{\mathbf{n}}_{\rm{q}}}}}{{\widehat {\mathbf{p}}}_n}} \right\}
\end{split}
\end{equation}

Applying ${{\mathbf{p}}_n}=\widehat {\mathbf{p}}_n^H+{{\boldsymbol{\delta }}_n}$, ${\left| {{{[{{\widehat {\mathbf{p}}}_n}]}_m}} \right|^2}\sim\Gamma (1,{\alpha _{n}}{\beta _{n}}{\left| {{\chi }} \right|^2})$ and ${\left| [{{{\boldsymbol{\delta }}_n}}]_{m}\right|}^2\sim\Gamma \left( {1,{{\left| \chi  \right|}^2}{\beta _n}(1 \!-\! {\alpha _n})} \right)$, we can obtain
\begin{equation}\label{eq:rate N simple}
{\mathbb{E}\left\{ {{{\left| {\widehat {\mathbf{p}}_n^H{{\mathbf{p}}_n}} \right|}^2}} \right\}}={{ {M^2 { {{\alpha^2_{n}}{\beta^2_n}{{\left| {{\chi }} \right|}^4}} } } }} + M { {{\alpha _{n}}\beta _n^2{{\left| {{\chi }} \right|}^4}} }.
\end{equation}
In a similar manner,  we can get 
\begin{equation}\label{eq:rate D1 simple}
\mathbb{E}\left\{ {\sum\limits_{k = 1,k \ne n}^K {{{\left| {\widehat {\mathbf{p}}_n^H{{\mathbf{p}}_k}} \right|}^2}} } \right\} = \sum\limits_{k = 1,k \ne n}^K { {\left(M {{\alpha _{n}}{\beta _k}{\beta _n}{{\left| {{\chi }} \right|}^4}} \right)} }.
\end{equation}
Applying ${\mathbf C}_{{\mathbf{n}}_{\rm{RF}}}\!=\!\sigma^2{\bf I}_{M}$ and ${\left| {{{[{{\widehat {\mathbf{p}}}_n}]}_m}} \right|^2}\!\sim\!\Gamma \left(1,{\left| \chi  \right|^2}\beta_n \alpha_n\right)$, we get
\begin{equation}\label{eq:rate D2 simple}
\mathbb{E}\left\{ \widehat {\mathbf{p}}_n^H{{\mathbf{C}}_{{{\mathbf{n}}_{{\rm{RF}}}}}}{{\widehat {\mathbf{p}}}_n} \right\}+ \mathbb{E}\left\{{{\left\| {{{\widehat {\mathbf{p}}}_n}} \right\|}^2} \right\} =M {{\left( {\sigma^2 + 1} \right){\alpha _{n}}{\beta _n}{{\left| {{\chi }} \right|}^2}} }.
\end{equation}
Applying ${{\mathbf{p}}_n}=\widehat {\mathbf{p}}_n^H+{{\boldsymbol{\delta }}_n}$, ${\left| {{{[{{\widehat {\mathbf{p}}}_n}]}_m}} \right|^2}\sim\Gamma (1,{\alpha _{n}}{\beta _{n}}{\left| {{\chi }} \right|^2})$, ${\left| [{{{\boldsymbol{\delta }}_n}}]_{m}\right|}^2\sim\Gamma \left( {1,{{\left| \chi  \right|}^2}{\beta _n}(1 \!-\! {\alpha _n})} \right)$ and ${{{\mathbf{C}}_{{{\mathbf{n}}_{\rm{q}}}}}}$ which is defined in \eqref{eq:covar of nq}, we get 
\begin{align}\label{eq:rate D3 simple}
&\mathbb{E}\left\{ {\widehat {\mathbf{p}}_n^H{{\mathbf{C}}_{{{\mathbf{n}}_{\rm{q}}}}}{{\widehat {\mathbf{p}}}_n}} \right\}=\eta (1 - \eta )M\left( {{\sigma ^2} + 1} \right){\alpha _n}{\beta _n}{\left| \chi  \right|^2}+\\
&\eta (1 - \eta ){\rho _{\rm{u}}}M{\alpha _n}{\beta _n}\left({\alpha _n}{\beta _n}{\left| \chi  \right|^4} + {\beta _n}{\left| \chi  \right|^4} + {\left| \chi  \right|^2}\sum\limits_{i = 1,i \ne n}^K  ( {{\beta _i}{{\left| \chi  \right|}^2}} )\right). \notag
\end{align}
Substituting \eqref{eq:rate D1 simple}, \eqref{eq:rate D2 simple} and \eqref{eq:rate D3 simple} into \eqref{eq:D temp}, and substituting \eqref{eq:rate N simple} into \eqref{eq:rate approx temp}, then we get
\begin{equation}
\widetilde{R}_n = {\log _2}{\left({1 + \frac{{\eta}{\rho_{\rm u}}{{\left| \chi  \right|}^2}{\beta _n}\left({{\alpha_n}M + 1} \right)}{\cal D}} \right)},
\end{equation}
where ${\cal D}$ is given by
\[
{\cal D}={\rho _{\rm{u}}}{\left| \chi  \right|^2}\left( {\sum\limits_i^K {\left( {{\beta _i}} \right)}  - \eta {\beta _n}} \right) + (1 - \eta ){\alpha _n}{\rho _{\rm{u}}}{\left| \chi  \right|^2}{\beta _n} + {\sigma ^2} + 1.
\]
\end{proof}
\end{NoHyper}
\bibliographystyle{IEEEtran}

\end{document}